\documentclass[amsmath,superscriptaddress,amssymb,preprint]{revtex4-1}
\usepackage{graphicx}
\usepackage[normalem]{ulem}
\usepackage{dcolumn}
\usepackage{xcolor}
\usepackage{pifont}
\usepackage{lineno}

\newcommand{\gst}{Ge$_2$Sb$_2$Te$_5$~}
\newcommand{\sbs}{Sb$_2$S$_3$~}
\newcommand{\st}{Sb$_2$Te$_3$~}
\newcommand{\sbsns}{Sb$_2$S$_3$}
\newcommand{\gstns}{Ge$_2$Sb$_2$Te$_5$}
\newcommand{\sn}{Si$_3$N$_4$~}
\newcommand{\snns}{Si$_3$N$_4$}

\begin{document}
 \title{Wide band gap phase change material tuned visible photonics}

\author{Weiling Dong}
\affiliation{Singapore University of Technology and Design, 8 Somapah Road, Singapore, 487372}
\author{Hailong Liu}
\affiliation{Singapore University of Technology and Design, 8 Somapah Road, Singapore, 487372}
\author{Jitendra K Behera}
\affiliation{Singapore University of Technology and Design, 8 Somapah Road, Singapore, 487372}
\author{Li Lu}
\affiliation{Singapore University of Technology and Design, 8 Somapah Road, Singapore, 487372}
\author{Ray J. H. Ng}
\affiliation{Singapore University of Technology and Design, 8 Somapah Road, Singapore, 487372}
\author{Kandammathe Valiyaveedu Sreekanth}
\affiliation{Division of Physics and Applied Physics, School of Physical and Mathematical Sciences, Nanyang Technological University, 21 Nanyang Link, Singapore, 637371}
\author{Xilin Zhou}
\affiliation{Singapore University of Technology and Design, 8 Somapah Road, Singapore, 487372}
\author{Joel K.W. Yang}
\affiliation{Singapore University of Technology and Design, 8 Somapah Road, Singapore, 487372}
\affiliation{Institute of Materials Research and Engineering, A$*$STAR, $\#$08-03, 2 Fusionopolis Way, Innovis, Singapore 138634}
\author{Robert E. Simpson}
\affiliation{Singapore University of Technology and Design, 8 Somapah Road, Singapore, 487372}
 \maketitle




\setlength\columnsep{6pt}
\textbf{Light strongly interacts with structures that are of a similar scale to its wavelength; typically nanoscale features for light in the visible spectrum.
However, the optical response of these nanostructures is usually fixed during the fabrication.
Phase change materials offer a way to tune the properties of these structures in nanoseconds.
Until now, phase change active photonics use materials that strongly absorb visible light, which limits their application in the visible spectrum. 
In contrast, Stibnite (\sbsns) is an under-explored phase change material with a band gap that can be tuned in the visible spectrum from 2.0 to 1.7 eV. 
We deliberately couple this tuneable band gap to an optical resonator such that it responds dramatically in the visible spectrum to \sbs reversible structural phase transitions. 
We show that this optical response can be triggered both optically and electrically.
High-speed reprogrammable \sbs based photonic devices, such as those reported here, are likely to have wide applications in future intelligent photonic systems, holographic displays, and micro-spectrometers.}

Fundamentally three mechanisms that can be used to tune the optical properties of nanostructures: (1) employing controllable refractive indices materials as surroundings; (2) mechanically changing the nanostructure shape and size; and (3) changing the charge density and dielectric function of the plasmonic materials themselves. 
For example, the tuneable refractive index of liquid crystals (LCs) has been used to change the spectral response of metamaterials\cite{kossyrev2005electric, liu2012optically, cetin2013thermal}. 
However, the small change in refractive index and the relatively slow switching speed, which is usually on a millisecond to microsecond scale, limit both the tuneability and modulation rate\cite{khoo2008theory}. 
The second approach relies on changing the separation distance, size, or geometry of plasmonic structures in the photonic resonator\cite{zhu2010mechanically, cataldi2014growing, millyard2012stretch, yamaguchi2013electrically, khirallah2015nanoelectromechanical}. 
The main problem with the second approach is mechanical stress, which tends to limit the switching endurance and the tuning speed. 
The third approach is achieved by varying the carrier density and dielectric function of the constituent plasmonic materials, such as metals\cite{novo2009electrochemical,rotenberg2009tunable}, semiconductors\cite{jain2013doped} and graphene\cite{gao2013excitation}.
However, the high intrinsic carriers density in metal and relatively weak plasmonic response in semiconductors result in a limited spectral response.

Phase change material (PCM) active photonics exploits refractive index switching materials. 
The prototypical phase change material, \gstns, can be reversibly switched on a sub-nanosecond time scale billions of times, and once switched no energy is needed to maintain the switched state\cite{loke2012breaking, simpson2011interfacial, waldecker2015time}.
This allows the photonics devices to be programmed to have a specific response.
\gst exhibits a substantial refractive index change when switched between different structural states. 
Indeed phase change switching in tellurium--based materials has been commercialised for optical and electrical data storage applications\cite{wuttig2007phase}. 
\gst has also been used to demonstrate switchable reflective colour display devices that are based on thin film Fabry-Perot cavities and plasmonic nanograting devices\cite{hosseini2014optoelectronic, yoo2016multicolor, dong2016wideband, wuttig2017phase}. 
However, using \gst to tune visible photonics devices is non-ideal due to the high absorption at above band gap photon energies and the relatively small change in the real component of the refractive index, $\Delta\textrm{Re(n)}$, at visible frequencies.
The small $\Delta\textrm{Re(n)}$ together with using ultra-thin \gst layers, which are necessary to minimise excessive absorption of the visible light\cite{hosseini2014optoelectronic}, further restrict the phase transition induced optical path length change.
Netherless, the potential of Ge-Sb-Te active photonics is apparent in the infrared, where its absorption coefficient is low, and $\Delta\textrm{Re(n)}$ is larger\cite{cao2016controlling, mkhitaryan2017tunable, michel2014reversible, tittl2015switchable, li2016reversible}. 
A further advantage of PCMs is that they maintain their structural state and only require energy during the switching process, which is a clear advantage over liquid crystals and mechanically tuned photonic devices.

For tuneable visible photonics, phase change materials with a low optical absorption in the visible spectrum are superior to highly absorbing materials, such as \gstns. 
The quality factor (Q-factor) of optical resonators is a measure of how much energy is lost from the optical cavity, and this determines the spectral purity.
Clearly, the imaginary component of the refractive index of tellurium--based PCMs deems these materials unsuitable for high-Q narrow band-pass filters in the visible spectrum. 
Moreover, for new tuneable transmissive filters, the transmission should be greater than 80\%\cite{Yun15}. 
To solve the absorption problem, new phase change materials with wider optical band gaps are required to tune photonic structures.
Until now, however, there are no reports of wide band-gap phase change material tuned photonics.

Herein, we avoid using tellurides such as \gst and propose an unconventional PCM, \sbsns, which has a much larger band gap and concomitantly lower absorption.
Generally, sulphides have a much wider band gap than selenides, and selenides have a larger band gap than tellurides.
Although sulphides' optical band gap is generally smaller than that of oxides, their higher refractive indices and low phonon frequencies make them attractive for applications that require high transmission from the visible to the mid-infrared\cite{eggleton2011chalcogenide}. 
The overall objective of this work is to demonstrate that wide band gap phase change materials with a tuneable absorption edge are highly suitable for active visible photonics.

Stibnite (\sbsns) is one of a number of chalcogenide PCMs that were studied in the 1990s for rewritable optical data storage\cite{arun1997effect}. 
However, the \sbs band gap is larger than the photon energy of the diode lasers that were used to read from and write to compact discs, which deemed \sbs unsuitable for optical data storage devices. 
Here we show that \sbs is misunderstood, and is well suited to both electrical and laser tuned active photonics. 
The reasons for this assertion are: 
(1) \sbs is a wide band gap PCM that exhibits a large absorption edge shift at visible frequencies,
(2) \sbs exhibits a large $\Delta\textrm{Re(n)}$ in the visible spectrum during the structural phase transition,
(3) Both the amorphous and crystalline phases of \sbs are stable at room temperature, which is important for non-volatile programming of optical devices. 
(4) The phase transition temperatures of \sbs are accessible by diode laser heating. 
(5) \sbs has a crystallisation activation energy ($\Delta$E$_{ac}$) of 2.0 eV (see Supplementary Information Figure S1), which is similar to that of \gst (2.3 eV) \cite{Tominaga2009jjap}.

\begin{figure*}
\centering
\includegraphics[width=0.60\textwidth]{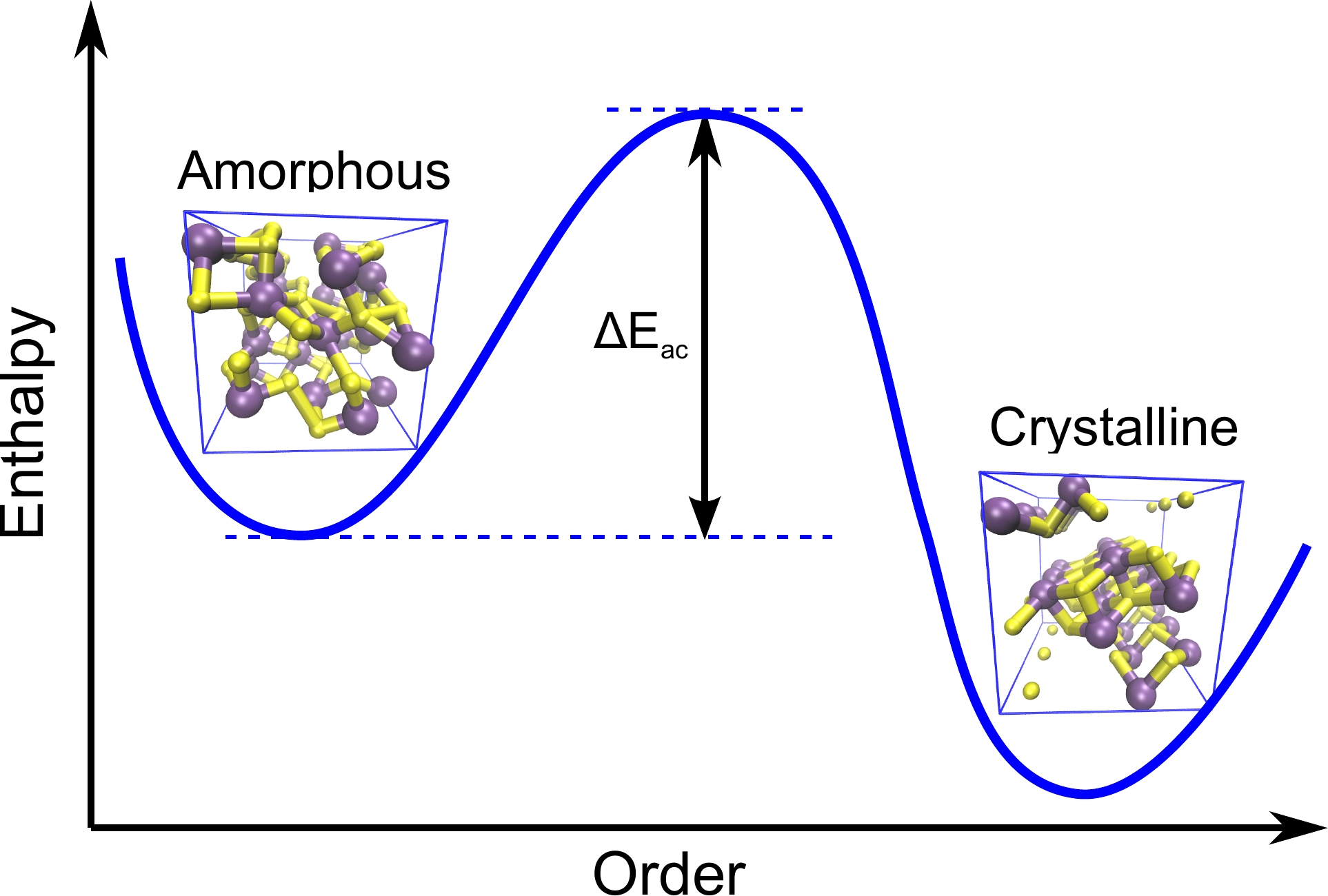}
\caption{\textbf{Atomic structures:} The amorphous and crystalline structures of \sbs are indicated on an enthalpy--order schematic plot.}
\label{atomic}
\end{figure*}

Herein, we demonstrate that when \sbs is used in resonator structures, it can easily be amorphised and reversibly switched on a nanosecond time scale between its amorphous and crystalline states.
Indeed, we find that the low absorption and fast switching characteristics of \sbs provide a substantial change to the resonant wavelength of reflective photonic devices in the visible spectrum.
This is in stark contrast to the study in the 1990s that concluded \sbs as a ``write once and read many times" data-storage material\cite{arun1999laser, arun1997laser}.
\sbs crystallises from an amorphous state into an orthorhombic structure at temperatures sufficient for its atoms to overcome the 2.0 eV energy barrier that separates the amorphous and crystalline states\cite{bayliss1972refinement}. 
This is depicted by the enthalpy--structural order diagram shown in Figure \ref{atomic}. 
The amorphous state is less stable than the crystalline structure, but the 2.0 eV activation energy barrier prevents it from spontaneously crystallising at room temperature. 
Crystallisation is achieved by heating \sbs to temperatures higher than 573~K, whilst amorphisation involves heating above its 823~K melting temperature\cite{Massalski90} and then rapidly quenching to `freeze-in' the disordered amorphous state.

\begin{figure*}
\centering
\includegraphics[width=0.5\textwidth]{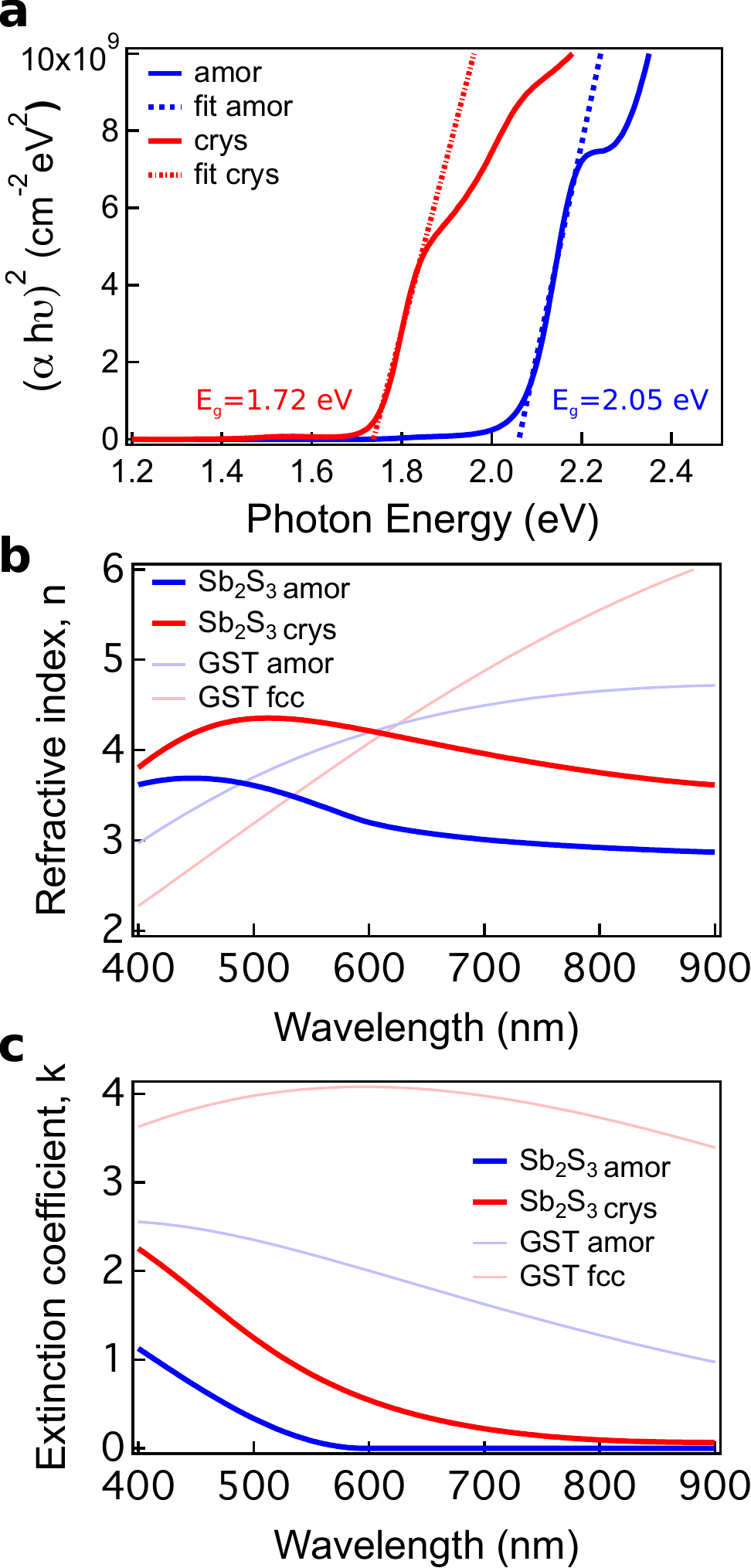}
\caption{\textbf{Optical properties of \sbs and \gstns:} \textbf{a}, Tauc plot showing the optical band gap of amorphous and crystalline \sbsns. 
The linear fit to the plot indicates that amorphous \sbs has a band gap of 2.05 eV, and crystalline \sbs has a band gap of 1.72 eV. 
\textbf{b}, The measured real component of refractive indices of \sbs and \gst in the amorphous and crystalline states. 
\textbf{c}, The measured imaginary component of refractive indices of \sbs and \gst in the amorphous and crystalline states.}
\label{nk}
\end{figure*}

We find that \sbs has very attractive properties for reprogrammable and tuneable visible photonics.
The absorption edges of \sbs in the amorphous and crystalline states are shown in Figure \ref{nk}a. 
Amorphous \sbs has a band gap of 2.05~$\pm$~0.05 eV, and crystalline \sbs has a band gap of 1.72~$\pm$~0.05 eV. 
This corresponds to a 115~nm absorption edge red-shift upon crystallisation. 
In comparison to \sbsns, \gst has a band gap of 0.5 and 0.7 eV in the crystalline and amorphous states\cite{lee2005investigation}, which leads to strong absorption at visible frequencies.
\gst has a much higher extinction coefficient than \sbs at visible frequencies, as shown in Figure \ref{nk}c.
The crystalline state of \sbs has a higher refractive index than the amorphous state across the visible spectrum from 400~nm to 900~nm, as shown in Figure \ref{nk}b. 
The maximum $\Delta$n $ \approx $ 1 is at 614~nm, whilst the extinction coefficient is close to zero for the amorphous state at wavelengths greater than 605~nm, and for the crystalline state at wavelengths greater than 721~nm. 
This is important because a phase transition in \sbs will change the absorption wavelength and also the concomitant condition for reflection and absorption of a thin film photonic resonator in the visible.
In comparison to \gstns, these characteristics of \sbs imply that it is a more suitable phase change material for non-volatile tuneable visible photonics. 

\begin{figure}[htbp] 
   \centering
   \includegraphics[width=0.75\textwidth]{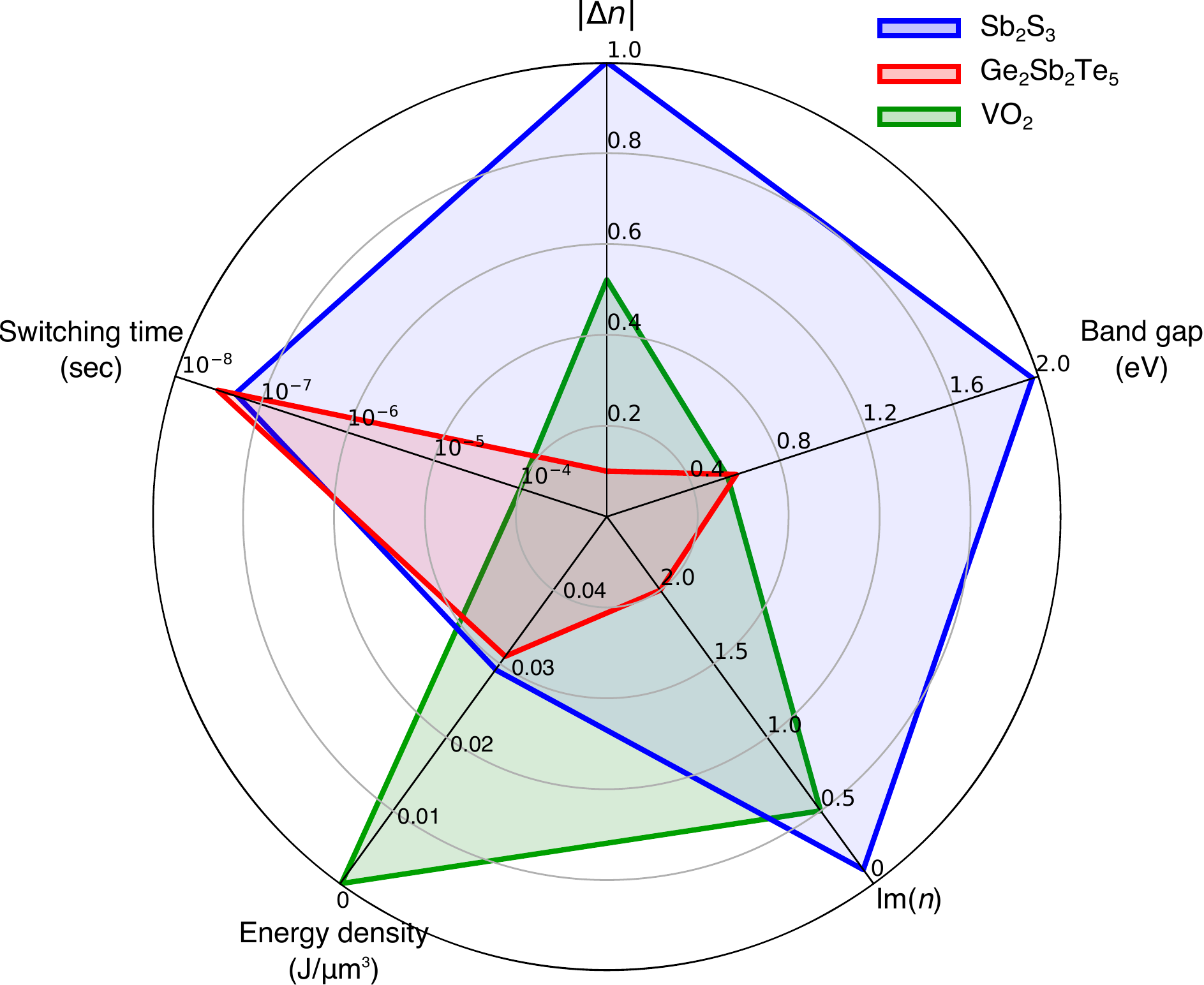} 
   \caption{A radar chart showing that the refractive index change, switching time, refractive index extinction coefficient ($\textrm{Im[n]}$), and band gap of \sbs are far superior to both \gst and VO$_2$ in the visible spectrum. 
   The optical constants were compared at 600 nm.
   The axes are arranged such that desirable properties are plotted at larger radii.
   The data is also presented in Table S1 in the Supplementary section.}
   \label{radar}
\end{figure}

In summary, \sbs shows a substantial change in refractive index, has a larger band gap and is far less absorbing than commonly used phase change materials, such as \gst and VO$_2$. 
We show later that the \sbs crystallisation switching time is similar to the \gst alloy, at 70 ns.
The switching energy density of VO$_2$ is lower than that of \sbsns, however, VO$_2$ requires a constant energy supply to hold its phase and optical properties.
In contrast \sbs exhibits a non-volatile change to its optical constants and only requires energy during the switching process.
The radar chart that illustrates the superior properties of \sbs relative to other tuneable media for active photonics devices in the visible spectrum is shown in Figure \ref{radar}.
The $|$$\Delta\textrm{n}$$|$ and $\textrm{Im(n)}$ of \sbsns, VO$_2$ and \gst are compared at a wavelength of 600 nm. 
A table that compares different tuning approaches to realise active photonic devices is shown in the Supplementary Information Table S1.


\begin{figure*}[!htbp]
\centering
\includegraphics[width=0.9\textwidth]{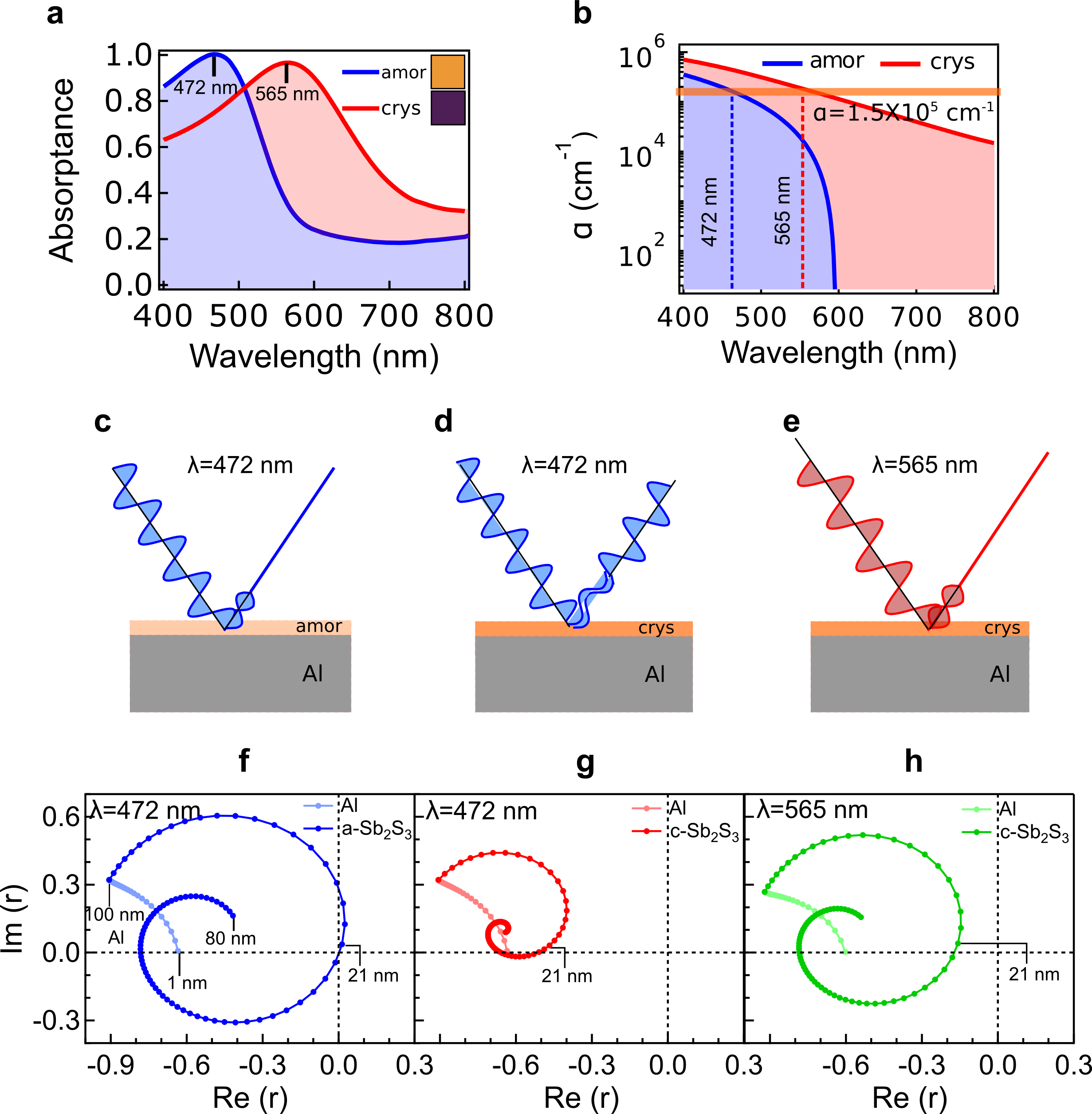}
\caption{ \textbf{\sbs optical changes:} \textbf{a,} Simulated absorptance spectra of 21~nm thick \sbs in amorphous and crystalline states on a 100~nm thick Al film. \textbf{b,} Calculated absorption coefficient of \sbsns. Schematic showing \textbf{c,} perfect absorbance of $\lambda$=472~nm light in the Al-\sbs structure in the amorphous and \textbf{d,} strong reflection of $\lambda$=472~ nm light in the crystalline state. \textbf{e,} The crystalline state shows strong reflection at a longer wavelength of $\lambda$=565~nm. \textbf{f,} The complex reflection coefficient of Al/amorphous \sbs structure as a function of thickness at $\lambda$=472~nm. The reflection coefficient is $r=0+0j$ when the thickness of amorphous \sbs is 21~nm. \textbf{g,} In the crystalline state at $\lambda$=472~nm, the structure reflection coefficient increases to $r$=-0.5+0$j$ and the strongly absorbing wavelength red-shifts to $r=-0.2+0j$, which is seen in \textbf{h}. }
\label{mechanism}
\end{figure*}

Here we show that actively tuning the damping coefficient of a critically damped PCM resonator by switching the \sbs absorption edge results in a radical spectral response at visible frequencies. 
We start with a simple structure with 21~nm \sbs on a 100~nm Al film. 
Perfect absorption of incident light is achieved by finely balancing the incident light reflected from the structure's first interface with the absorption of light that is transmitted into the structure and then reflected out from buried interfaces.
Figure \ref{mechanism}a shows that this structure perfectly absorbs at a wavelength, $\lambda=472$~nm.
When \sbs is crystallised, this absorptance maximum switches to $\lambda=565$~nm.
Perfect absorption in these structures occurs when the intrinsic absorption of the component layers reduces the light intensity reflected from buried interfaces to balance the light reflected from the top air-\sbs interface. 
In addition, any optical phase changes that result from path length differences and interfacial reflections must also cancel. 
Thin sub-$\lambda$/4 absorbing semiconductor films form strong resonant behaviour at visible wavelengths due to a non-$\pi$ phase shift at the semiconductor--metal interface\cite{kats2013nanometre}.
Here, this effect can cause destructive interference in the \sbs films despite their thickness being an order of magnitude thinner than the wavelength of light.

The necessary balance of absorption and reflection is clear in Figure \ref{mechanism}b.
It shows the absorption coefficient of amorphous and crystalline \sbs as a function of wavelength.
The absorption coefficient was calculated from the extinction coefficient shown in Figure \ref{nk}c.
We see that the amorphous and crystalline absorptance maxima occur at wavelengths of 472~nm and 565~nm, where the absorption coefficient of \sbs is 1.5$\times$10$^{5}$~cm$^{-1}$.
In effect, the wavelength for perfect absorption must increase across the absorption edge in order to balance the light reflected from the top air-\sbs interface. 
Despite the difference between the real component at $\lambda$=472~nm and $\lambda$=565~nm being $\Delta\textrm{Re(n)}$=0.3, the PCM film is just 21~nm thick, and the change in optical path length between the amorphous and crystalline states is minimal. 
The significant red-shift to the \sbs absorption-edge in the visible spectrum, which is seen in Figure \ref{nk}c and Figure \ref{mechanism}b, dominates the interference condition. 

A further way to see the effect of the edge shift is to plot the complex reflection coefficient, $r$, as a function of the \sbs film thickness at the absorptance maximum wavelengths, $\lambda$=472~nm and $\lambda$=565~nm. 
Figures \ref{mechanism}f-h show how the reflection coefficient for an Al/\sbs structure changes as the structure's thickness increases. 
Initially, the Al and \sbs are not present, and the reflection coefficient only has a real component, $r$=-0.6, which gives rise to the 36\% reflectivity of the silicon substrate.
As the aluminium thickness increases, the reflection coefficient turns away from the real axis due to a small amount of absorption.
Once the Al is 100~nm thick, the \sbs layer is added and its thickness increases. 
As can be seen in Figure \ref{mechanism}b, \sbs has an absorption coefficient of 1.5$\times$10$^{5}$~$cm^{-1}$ for the amorphous and crystalline states at $\lambda=472$~nm and $\lambda=565$~nm respectively, hence the imaginary component of the reflection coefficient, $\textrm{Im(r)}$, is large.
When its thickness is 21~nm, and the total structure thickness is 121~nm, the reflectance coefficient is $r$=$0+0j$, and the material is perfectly absorbing at $\lambda$=472~nm. 
Further increasing the \sbs thickness increases the net loss of the system, and the reflected intensity from the top air-\sbs interface is no longer suppressed.
Hence the reflectance coefficient moves away from the origin.
For comparison, similar plots for \gstns, which is highly absorbing in the visible, are given in Supplementary Figure S2.

Crystallising the \sbs increases the loss and therefore produces a much smaller spiral radius in the complex plane at $\lambda$=472~nm, as shown in Figure \ref{mechanism}g.
It is now impossible at this wavelength to produce perfect absorption by simply increasing the \sbs thickness.
However, increasing the incident wavelength to $\lambda$=565~nm decreases the absorption, and the radius of the complex plane spiral increases such that the reflectance coefficient of a 21~nm thick \sbs film is close to 0, see Figure \ref{mechanism}h. 
Indeed, we observe a peak in the absorptance spectrum in Figure \ref{mechanism}a.
Note that for \gst films, the perfect absorption peaks occur in the infrared and are not present in the visible spectrum unless additional dispersive elements are added to the structure \cite{dong2016wideband}.
We conclude that the imaginary component of the \sbs refractive index, $\textrm{Im(n)}$, predominately drives the change in absorptance maximum.
Thus, this design mainly exploits the change in $\textrm{Im(n)}$ rather than $\textrm{Re(n)}$, which is usually used to tune \gst devices\cite{wuttig2017phase, tittl2015switchable, li2016reversible}.

\begin{figure*}[!htbp]
\centering
\includegraphics[width=1\textwidth]{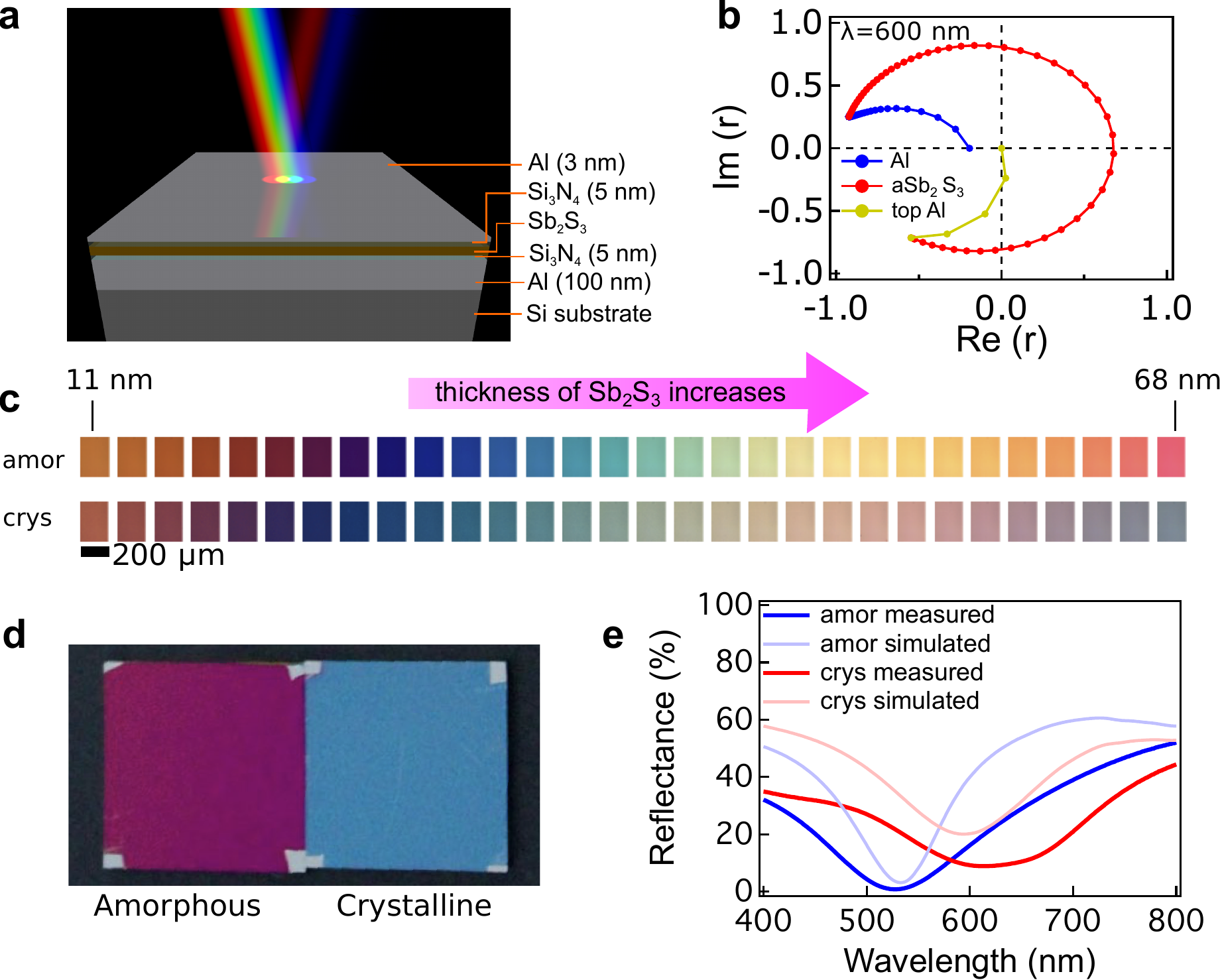}
\caption{\textbf{Reflectance:} \textbf{a,} Schematic structure of the fabricated sample. \textbf{b,} Complex reflection coefficient at $\lambda=600$ nm for a structure composed of a 100~nm thick Al layer,  a 51~nm thick amorphous \sbs layer, and a 4~nm thick top layer of Al. \textbf{c,} Achievable colours of the structure shown in (a) with the thicknesses of \sbs ranging from 11~nm to 68~nm, with \sbs in amorphous and crystalline states. \textbf{d,} Photograph of the structure shown in (a) with the 24~nm thick \sbs layer in the amorphous and crystalline states. \textbf{e,} Experimentally measured and simulated reflectance spectra of the structure shown in (a) with 24~nm thick \sbs layer in the amorphous and crystalline states. }\label{reflectance}
\end{figure*}

To further show the effect of using the \sbs absorption edge change to tune at visible wavelength, an Al/\snns/\sbsns/\snns/Al stacked structure was designed.
This simple device structure was deliberately chosen to demonstrate the superior performance of \sbsns, rather than a new type of photonics device. 
The bottom and top Al thicknesses were 100~nm and 3~nm, respectively. 
The device structure is schematically shown in Figure \ref{reflectance}a.
\sn diffusion barriers was also included between the Al and \sbs layers to prevent inter-layer diffusion, which is known to be a problem when metals interface directly with phase change chalcogenides\cite{chew2017chalcogenide}. 
The top Al layer was included because it increases the Q-factor of the resonator and also allows the PCM to be electrically addressed, see Figure \ref{reflectance}b and Supplementary Figure S3.
The thickness of the \sbs layer is varied from 11~nm to 67~nm to shift the absorptance maximum wavelength. 

The reflectance spectra of the Al/\snns/\sbsns/\snns/Al structures were measured and simulated, see Supplementary Figure S4.
Effectively, \sbs switches from a low-loss to lossy dielectric as the wavelength is scanned from the near-infrared to the short-visible.
The \sbs structural phase transition red-shifts the resonator's natural frequency, and this causes the strong reflected colour change, which is clearly seen in the photographs of the sample that are displayed in Figure \ref{reflectance}c and Supplementary Figure S5.
Figures \ref{reflectance}d-\ref{reflectance}e show the photographs and corresponding reflectance spectra for the sample in the amorphous and crystalline states when the thickness of \sbs is 24~nm.
For the 24~nm thick \sbs structure shown in Figure \ref{reflectance}e the absorptance peak is red-shifted by 110~nm from 520~nm to 630~nm. 
We also see in Supplementary Figure S4 that as the thickness of the \sbs layer increases, the peak absorptance wavelength is red-shifted across the visible spectrum from 425~nm to 710~nm.
The corresponding simulations reproduce the key spectral features that are observed in the measurement.

The red-shift to the spectra shown in Figures \ref{reflectance}e is directly due to the intrinsic absorption edge of \sbs narrowing from 2.0 eV to 1.7 eV.
The metal--\sbsns--metal structure is designed to resonate at wavelengths close to this absorption-edge, which amplifies the effect of the absorption edge switching on- and off- the resonant wavelength, thus producing the large change in reflected colour.
However, crystalline \sbs is slightly more absorbing than the amorphous state, and this causes a broader resonance, as shown in Figure \ref{reflectance}e and Supplementary Figures S4--S5.


Achieving a large colour shift from a 24~nm thick \sbs film may seem surprising considering the change in the $\Delta\textrm{Re(n)}$ $\approx$ 0.9 produces a 32~nm optical path length change at 550~nm. 
This corresponds to a phase shift of $\phi$=0.36~rad, which has a negligible effect on the reflected colour. 
However, for this metal-\sbs structure, the colour change is dominated by the \sbs absorption edge red-shift effect on the imaginary component of the refractive index, which at $\lambda$=550~nm changes from 0.08 to 0.83, see Figure \ref{nk}c.

\begin{figure*}[!htbp]
\centering
\includegraphics[width=0.9\textwidth]{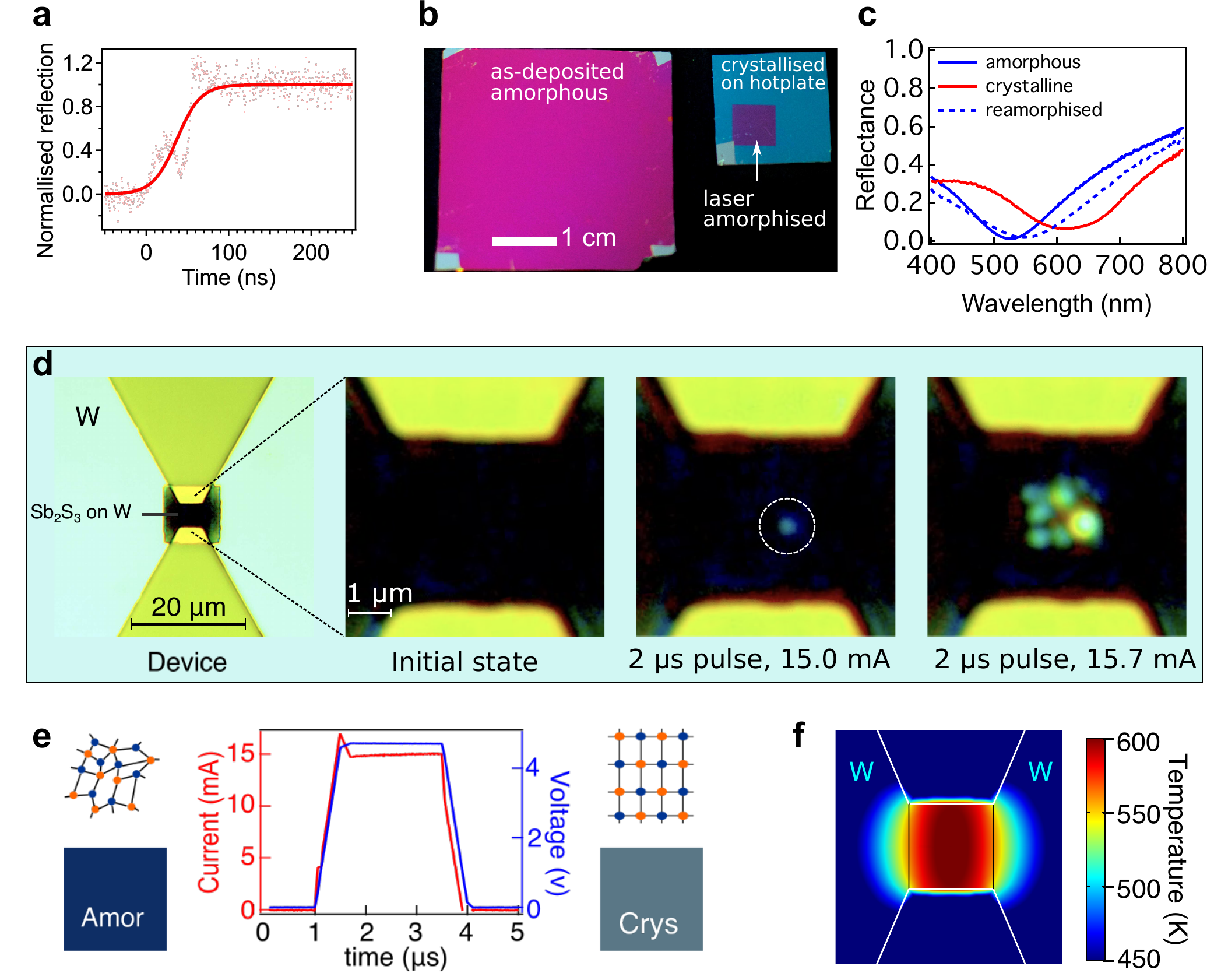}
\caption{\textbf{Switching demonstration:} \textbf{a,} Normalised reflection versus time during laser crystallisation of \sbsns. \textbf{b,} Optical photograph of the Al/\sbsns/Al structure with \sbs thickness of 20~nm, with \sbs in the as-deposited amorphous, crystalline, reamorphised states. \textbf{c,} Reflectance spectra of the Al/\sbsns/Al structure with \sbs thickness of 20~nm, with \sbs in the as-deposited amorphous, crystalline, reamorphised states. \textbf{d,} Optical photograph of an electrically switchable colour filter device and zoomed-in optical photographs of a device before applying a pulse, after a 2~$\mu$s,~15~$\pm$~1~mA pulse, and after a 2~$\mu$s,~15.7~$\pm$~1~mA pulse. \textbf{e,} Schematic structure and simulated colour of W/amorphous \sbsns; current versus time and voltage versus time of electrical pulse applied to W/\sbsns; schematic structure and simulated colour of crystalline state of W/crystalline \sbsns. \textbf{f,} The temperature distribution on the surface of W computed using a time dependent Joule heating-thermal diffusion finite element model. }
\label{switch}
\end{figure*}

Active photonics devices must be tuneable or capable of switching between different states. 
Here we demonstrate both reversible laser switching and electrical switching.
The dichotomy for well-studied phase change materials along the GeTe-\st tie-line is that they are stable at room temperature but quickly and efficiently undergo a structural phase transition at slightly elevated temperatures\cite{wuttig2007phase}.
This is also a characteristic of \sbsns. 
Here we show that \sbs can also be crystallised and amorphised at a similar speed to \gstns. 
A pump-probe nanosecond laser system was used to simultaneously heat the \sbs film and measure the change in reflectivity of the heated sample\cite{behera2017laser}. 
Figure \ref{switch}a shows the normalised reflectivity as a function of time for the crystallisation of \sbs in the Al(100~nm)/\snns(5~nm)/\sbsns(19~nm)/\snns(5~nm)/Al(3~nm) structure. 
The amorphous film completely crystallises in 78~$\pm$~4~ns. 
These pump laser amorphised marks have a diameter of $\sim$1.5 $\mu$m.
Therefore, we rastered a 5~ns, $\lambda$=532~nm amorphising laser pulse over a 0.25 $cm^2$ area to make the amorphous region visible. 
A photograph of the amorphised area is shown in Figure \ref{switch}b.
The initial amorphous colour was pink and after crystallisation the colour changed to blue. 
We can clearly see that after rastering the amorphising laser, the structure's colour is switched reversibly back from the crystallised state to the amorphous state. 
Figure \ref{switch}c shows the corresponding reflectance spectra from the structure with \sbs in the as-deposited amorphous state, crystallised state, and the re-amorphised state.
A video showing the recrystallisation process is included in the Supplementary Information.
Note that the reamorphised state's spectra and colour are marginally less reflective, and this is due to ablation at the centre of the laser's Gaussian intensity profile.
We conclude that \sbs can be reamorphised in 5~ns to produce large reversible changes to the visible spectrum.
This is important because until now, it has been claimed that \sbs is a ``write once, read many times'' material\cite{arun1999laser, arun1997laser}, and the optical change is insignificant. 
However, by designing the resonant wavelength of \sbsns-metal structures to coincide with the absorption edge of \sbsns, strong interference effects can be exploited to efficiently absorb the 532~nm light in \sbs layers that are just 21 nm thick, which in turn allows amorphisation. 
We have included further data for amorphisation of other metal--\sbs structures in Supplementary section 4.


Many active photonics devices are electrically addressed, and therefore we show that these metal/\sbs filters can be electrically switched.
We adopted a line-cell design similar to those proposed for reconfigurable GHz switches\cite{champlain2016examination} but without the RF waveguide. 
The \sbs was deposited directly on top of a tungsten filament.
This configuration allows direct optical probing of the active \sbs area before and after applying electrical pulses.
The electric current through the W filament causes Joule heating. 
The heat is conducted into the \sbs layer, which provides the energy for it to undergo a phase transition between the amorphous and crystalline structural states.
Figure \ref{switch}d shows images of the device in the initial state before applying an electrical pulse, after applying a 2 $\mu$s, 15.0 $\pm$~1~mA pulse, and finally after applying a 2 $\mu$s, 15.7~$\pm$~1 mA pulse. 
We designed the structure to be highly absorbing in the visible spectrum when \sbs is amorphous, and consequently the image of the active area is dark.
The simulated colour for the W/\sbs structure in the amorphous and crystalline states is inset into Figure \ref{switch}e, whilst the spectra is shown in Supplementary Information Fig. S7.
The crystallised region should become a light blue-grey colour.
Applying the electrical pulse heats the amorphous structure, allowing it to crystallise at temperatures higher than 585~K.
An electrical Joule heating finite element simulation shows the temperature distribution in the \sbs film after applying a 2 $\mu$s, 15 mA pulse, see Figure \ref{switch}f.
This results in crystallisation at the hottest part of the active area, which according to the Joule heating finite element simulation temperature distribution, is the central area of the \sbs where the temperature is higher than the crystallisation temperature. 
Indeed, Figure \ref{switch}d illustrates that the 2 $\mu$s, 15~mA pulse produces a light blue-grey crystalline mark at the centre area of the device. 
This indicates that \sbs is readily switched using the electrical Joule heating device. 
The crystallised area can be increased by adjusting the electrical pulses applied to heat the device and optimising its thermal design. 
However, this particular device design is susceptible to small increases in the electrical current, which causes ablation of the hottest area in the sample. 
This is seen in Figure \ref{switch}d when a 15.7~mA, 2 $\mu$s pulse is applied to the sample. 

In conclusion, the wide band gap phase change materials \sbs has excellent properties for visible photonics and is therefore preferred over more conventional PCMs, such as \gstns, which was originally developed for data storage applications.
The bandgap of \sbs is large and produces an absorption edge which can be shifted in the visible spectrum in nanosecond time scales by structural phase transitions.
In contrast to exploiting a change in the real part of the PCMs refractive index, which is usually used to tune \gst infrared photonic resonators, we show that the \sbs band gap tuneability can be used to tune resonant photonics devices, thus producing a substantial change to the spectral response and resultant colour. 
The fast switching, large optical band gap, and tuneable refractive index deem \sbs a versatile phase change material for tuneable photonics at visible frequencies. 
Considering the lack of reported large band-gap phase change materials and the emergence of smart photonics, where flexibility and programmability of components and materials are required, wide band gap phase change materials, such as \sbsns, will lead to a broad range of innovative photonic devices.

\noindent
\section*{Experimental Section}

\textbf{Fabrication}
Aluminium films were deposited on Si(001) substrates by DC sputtering, and \sn and \sbs were deposited by radio frequency (RF) sputtering. 
The chamber base pressure was 4$\times$10$^{-5}$ Pa and the sputtering pressure was 0.5 Pa. 
The deposition rate was 0.5 �\AA/min from an \sbs alloy target with a diameter of 50.8 mm and a purity of 99.9 \%. 
\sn was deposited by using a gas mixture of Ar: N$_{2}$=8:2 from a Si target. 
To crystallise the \sbs in the optical filter, the samples were heated at 300 $^\circ$C for 30 minutes on a hot plate.\\
\textbf{Optical band gap measurement}
The optical band gaps were calculated by Tauc analyses\cite{tauc1968optical}.
As \sbs has a direct energy band gap, the $\alpha(\hbar\nu)\sim \hbar\nu$ - $E_g^2$ was used, where $\alpha$ is absorption coefficient, $\hbar\nu$~ is photon energy, $E_g$ is band gap\cite{jellison1996parameterization}.\\
\textbf{Ellipsometry}
The optical constants of \sbs in the amorphous and crystalline states were measured using an ellipsometry spectrometer (WVASE, J.A Woollam Co.). 
37~nm of as-deposited \sbs deposited on Si (001) substrate was used for measurement. 
A further sample with the same thickness was crystallised in a tube furnace at 300 $^\circ$C in an argon atmosphere for 20 minutes.
A heating rate of 4 K/min was used.
Both samples were measured by ellipsometry. 
The wavelength range of incident light is from 250~nm to 900~nm and the angle of incident light was 65.1$^{\circ}$.
The data was fitted by Lorentz and Tauc-Lorentz oscillator models for both amorphous and crystalline states. 
The fit parameters are listed in the Supplementary Information table S2, whilst the spectroscopic ellipsometric parameters and the computed refractive indices are available at www.actalab.com.\\
\textbf{Reflectance spectra}
Reflectance spectra of the Al/\snns/\sbsns/\snns/Al structures were measured using a QDI 2010 UV-visible-NIR range microspectrophotometer (CRAIC Technology Inc., California, USA). 
The reflectance spectra were normalised to an Al mirror, which was considered 100\% reflective across the measured spectral range. 
The absorptance maximum wavelengths were obtained by Gaussian fitting at the dip of the reflectance spectra. \\
\textbf{Laser switching}
A pump-probe system with pulse length from 20 ns to 2000 ns and pulse power up to 40 mW was used to switch the Al/\sbsns/Al structure\cite{behera2017laser}. 
A 658~nm pump laser with higher power was used to induce a phase transition in the \sbs film. 
To measure the change in reflected signal, a 100 $\mu$W and 1 $\mu$s probe laser was used.
A 635~nm probe laser was used to focus the probe beam on the write mark. 
The reflected probe signal was detected simultaneously by the fast silicon photodetectors.
The final data was measured by a digitising oscilloscope (NI PXIe-5162, 10 bit, 1.5 GHz).
A camera was used to image the sample surface. 
We assumed that the crystallised fraction was directly proportional to the change in the probe light.\\
\textbf{Optical microscope images}
To compare the colours of the optical filters with different \sbs thickness, the bright field optical micro images of the colour filter were taken using an Olympus BX51 microscope through a �10$\times$ objective lens. 
Each colour square is 680 $\mu$m $\times$ 680 $\mu$m. 

\section*{Acknowledgement}
This research was performed under the auspices of the SUTD-MIT International Design Center (IDC). 
The research project was funded by the Samsung GRO, the A-star Singapore-China joint research program Grant No. 1420200046, and the SUTD Digital Manufacturing and Design Centre (DManD) Grant No. RGDM 1530302.
We are grateful for fruitful discussions with Seokho Yun.

\vspace{1cm}
\small

\bibliographystyle{advancedmaterials}

\bibliography{Ref_Sb2S3_nature}


\end{document}